\documentclass[pre,aps,nofootinbib,twocolumn,showpacs]{revtex4}
\usepackage{graphics}
\usepackage{epsfig}

\begin{document}

\title{Crossovers from parity conserving to directed percolation
universality}

\author{G\'eza \'Odor (1) and N\'ora Menyh\'ard (2)}

\affiliation{(1) Research Institute for Technical Physics and
  Materials Science, \\
(2) Research Institute for Solid State Physics,
H-1525 Budapest, P. O. Box 49, Hungary}    

\begin{abstract}
The crossover behavior of various models exhibiting phase transition
to absorbing phase with parity conserving class has been 
investigated by numerical simulations and cluster mean-field method. 
In case of models exhibiting $Z_2$ symmetric absorbing phases 
(the NEKIMCA and Grassberger's A stochastic cellular automaton) 
the introduction of an external symmetry breaking field causes a 
crossover to kink parity conserving models characterized by 
dynamical scaling of the directed percolation (DP) and
the crossover exponent: $1/\phi\simeq 0.53(2)$. 
In case an even offspringed branching and annihilating random 
walk model (dual to NEKIMCA) the introduction of spontaneous 
particle decay destroys the parity conservation and results 
in a crossover to the DP class characterized by the
crossover exponent: $1/\phi\simeq 0.205(5)$. The two different 
kinds of crossover operators can't be mapped onto each 
other and the resulting models show a diversity within the 
DP universality class in one dimension. These 'sub-classes'
differ in cluster scaling exponents. 
\end{abstract}
\maketitle

\section{Introduction}

The study of nonequilibrium phase transitions is an important 
task of statistical physics. Genuinely nonequilibrium transitions can be
observed most easily in models exhibiting transition from an active
to an ``absorbing'' state, where the fluctuations are negligible, 
hence no return is possible.
The exploration of critical phenomena and universality classes of
simple, one component models has been started
\cite{DickMar,Hayeof,Orev,Svenrev} and important steps towards a full 
classification have been done \cite{EK06,Obook08}. 

For a long time it was a common belief that all continuous, 
nonequilibrium phase transitions belong to class of the directed 
percolation (DP) and a hypothesis advanced by Janssen and 
Grassberger \cite{DPuni,DPuni2,DPuni3}. This states that
in one component systems exhibiting continuous phase
transitions to a single absorbing state (without extra symmetry and
inhomogeneity or disorder) short ranged interactions can generate DP
class transition only. Despite the robustness of this class 
experimental observation is rare, owing to the high sensitivity to
disorder and long-range interactions. A very recent experimental 
study \cite{TKCS07} has reported clear and comprehensive 
experimental evidence of DP criticality.  
Later it was discovered that in systems with infinitely many 
frozen absorbing states  (IM type) \cite{PCP,PCP2,dimerr,TTP} 
like in the pair contact process (PCP) the static exponents 
coincide with those of the DP. 
The dynamical cluster spreading behavior is different owing to the 
long time memory generated by the frozen monomers \cite{OMSM98}. 
In the literature the transition type of PCP is often called DP type.
Very recently \cite{PP07} have investigated in $1+1$ dimension 
the crossover from PCP to DP type of models by introducing 
different absorbing state reduction mechanisms and nontrivial
exponents have been found. This finding confirms that 
strictly speaking the universality class of PCP and DP is
different.

The first example for clearly non-DP critical behavior was 
found among stochastic cellular automata (SCA) by Grassberger \cite{Gras84}. 
These models exhibit $Z_2$ symmetric absorbing states and an effective
kink dynamics, which follows an even offspringed branching and annihilating 
random walk (BARWe). In reaction-diffusion (RD) particle models, with 
BARWe dynamics ($A\to 3A$, $2A\to\emptyset$) such phase transition 
class was discovered by \cite{Taka,Jensen}. Since this is different 
from odd offspringed branching and annihilating random walks, 
in which DP class transition occurs, the name ``parity 
conserving'' (PC) was introduced to denote this class 
\footnote{However other different names for this class
like ``Directed Ising''(DI)  or ``Generalized voter (GV)'' 
or ``BARW'' can also be found in the literature}. 

An important example of the PC class behavior was discovered
in one-dimensional kinetic Ising models with combined 
zero temperature spin-flip and finite temperature spin-exchange
dynamics (NEKIM) \cite{Men94}. Here the domain walls between up 
and down spins follow BARWe dynamics and an exact duality 
transformation in one dimension between the NEKIM and the BARWe 
particle model was established by \cite{Mussa} \footnote{Note that
such mapping is not possible in higher dimensions, but since the
upper critical dimension is $d_c=4/3$ it is not important.}. 
Naturally the NEKIM exhibits two, $Z_2$ symmetric absorbing states.
This universality class has not been observed in nature yet,
however we pointed out \cite{qpccikk} that this model is very
insensitive to the quenched disorder, therefore it is a 
good candidate for experimental verification. 
In fact it's inactive phase, where annihilating random walk 
dominates has been observed \cite{1dexp} unless very strong
disorder fractures the medium. 

The introduction of a symmetry breaking external magnetic field, 
in NEKIM, which favors one of the absorbing states (but preserves the BARWe 
dynamics) was shown to change the type of transition from PC to DP 
type \cite{meod96}. Such crossover mechanism 
has been observed in other $Z_2$ symmetric models as well 
\cite{Parkh,Bas,Hin97,HKPP98,KHP99}.
On the other hand simulations \cite{KP95} and field theory 
\cite{Cardy-Tauber,CCDDM05} proved that a PC breaking 
$A\to\emptyset$ reaction in the BARWe model should also change 
the type of phase transition from PC to DP type. 

The more detailed study of crossover behavior among nonequilibrium
universality classes has been intensified in the recent years.
The universal crossover exponent, defined by the shape of
phase boundary ($r_c(w)$) as one introduces a relevant scaling 
field ($w$) (see for example \cite{LS84}),
\begin{equation} \label{crossscal}
r_c(w) \sim w^{1/\phi}
\end{equation}
has been determined in case of DP to the compact DP class
\cite{MDH96,Jans05,Svencross,DOS08}. Crossover between DP 
and isotropic percolation was investigated by field theory 
\cite{FTS94,FN97,JaStecross} and simulations \cite{Irescross}. 
Very recently the numerical exploration of the crossover behavior 
from the diffusive pair contact process (PCPD) to DP \cite{PP06}
has strengthened the existence of the independent PCPD class theory. 
In IM type models 'nontrivial crossovers to DP class' have been 
found \cite{PP07}. The crossover behavior from DP to mean-field,
generated by long-range diffusion, following the early studies
\cite{boccikk,OSzab94}, has currently been re-examined via numerical 
techniques \cite{DOS07} and field theory \cite{MH08} and  
estimates for the exponent $\phi$ have been provided.
Similar, diffusion driven crossover in case of PC class
was determined within the framework of NEKIM \cite{meorcikk}
long ago.

In the present paper we determine the crossover exponent from
PC to DP class in a SCA version of the NEKIM model \cite{meod96} 
(NEKIMCA) and compare it with other realizations of the PC class. 
In particular we confirm the universality of $\phi$ in case of 
$Z_2$ symmetry breaking fields by simulating Grassberger's A model.
We compare this crossover behavior with the outcome of 
parity conservation breaking in a BARWe model.

\section{Crossover of NEKIMCA in an external field}

The NEKIM exhibiting PC class transition was suggested by 
\cite{Men94} as a generalization of the Glauber Ising model 
\cite{gla63}. It is defined by the alternating application of 
a $T=0$ spin-flip sweep and a $T>0$ Kawasaki spin-exchange 
update of a one dimensional lattice. 
While the spin-flip dynamics generates annihilating random walk 
of kinks, the spin-exchange introduces a parity conserving 
branching ($A\to 3$) of the domain walls. Tuning the relative 
strengths of the reactions one can get a phase transition 
from an active to a kink-free, adsorbing state 
(the order parameter is the density of kinks).

It was realized in \cite{meod96} that branching reactions
appear automatically in the SCA version of the NEKIM, if the
spin-flip update is done synchronously due to overlaps. 
With this dynamics we can obtain a very simple model,
with PC type of critically, which can be implemented on a 
computer by efficient bit-coding.
The NEKIMCA spin updates, represented by bit field operations of
a computer word, generating BARWe reactions of the kinks
are the followings:
\begin{itemize}
\item Random walk of domain walls ($\bullet$): \\ 
$\uparrow \ \ \uparrow\bullet\downarrow \ \stackrel{w_i}{\longrightarrow} \
  \uparrow\bullet\downarrow \ \ \downarrow$ \\
generated by a spin-flip between oppositely oriented spins, with probability
$w_i$,
\item Annihilation of a pair of kinks: \\
$\uparrow\bullet\downarrow\bullet\uparrow \
  \stackrel{w_o}{\longrightarrow} \   \uparrow \ \ \uparrow \ \ \uparrow$ \\
generated by a spin-flip between a pair of same oriented spins, with
probability $w_o$,
\item Branching of a kink: \\
$\uparrow \ \ \uparrow\bullet\downarrow \ \ \downarrow   \  
\stackrel{w_i^2}{\longrightarrow} \  \uparrow\bullet\downarrow\bullet\uparrow\bullet\downarrow$ \\
can be generated by two overlapping spin-flips (with the probability $w_i^2$)
such that it causes a spin-exchange.
\end{itemize}
There are two independent parameters related to the
original parametrization of Glauber:
\begin{eqnarray}
w_i=\Gamma (1-\tilde\delta)/2 \\ w_o=\Gamma (1+\tilde\delta) \ .
\end{eqnarray}
In the present study we investigated the $\Gamma=0.35$ and  $\Gamma=2$
cases. The corresponding PC critical points, without external field, 
are located at $\tilde\delta_c=-0.535$ and $\tilde\delta_c=-0.416$ 
respectively.

\begin{figure}[ht]
\begin{center}
\epsfxsize=70mm
\epsffile{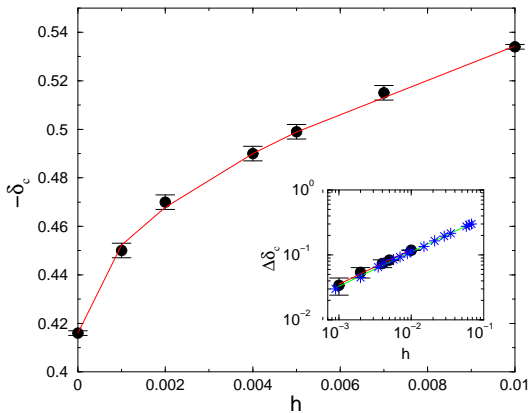}
\caption{(Color online) Critical point shift in the NEKIMCA model 
as a result of the external $h$ field for $\Gamma=2$. 
The solid line is power-law fitting on the numerical data points. 
The insert shows the scaling of the shift of $\tilde\delta$ 
for $\Gamma=2$ (bullets) and for $\Gamma=0.35$ (stars) on a 
log-log scale.} \label{NEKIMAfig}
\end{center}
\end{figure}

A single Monte Carlo step (MCS) consists of updating all sites at once 
as described in \cite{MeOdof} (throughout the paper the time is 
measured by MCS).
The simulations were reformed on $L=40 000$ sized lattices, with periodic
boundary conditions up to $t_{max}=10^6$ MCS. 
The change of critical point $\delta_c(h)$ has been determined for several
values of external field. The transition probabilities are modified
in the presence of an external magnetic field $H$ as:
\begin{eqnarray}
w_{i}^h = w_{i}(1 - h s_i), \\
w_{o}^h = w_{o}(1 - h s_i), \\
h = th({H\over kT}) \ .
\end{eqnarray}
Fig.~\ref{NEKIMAfig} shows the phase diagram in the $(h,\tilde\delta)$ 
plane, which is similar to the one we obtained in \cite{meod96} 
by simulations and cluster mean-field technique.
We have applied only random initial state simulations to find 
the points of the line of phase transition exhibiting kink density 
decay exponent $\alpha=0.1595$ of the DP \cite{Orev}. 
A clear power-law scaling of the critical point shift 
$\Delta\tilde\delta$ as the function of $h$ can be fitted with 
the form (\ref{crossscal}) in the region $0<h<0.01$ with 
$1/\phi=0.52(3)$ crossover exponent. The error-bars on Fig.~\ref{NEKIMAfig}
come from our numerical estimates for the critical point shifts
and the least squares error estimate fitting procedure.
This value is close to the early numerical estimates:
$\phi=2.1(1)$ by \cite{Bas} and $\phi=2.24(10)$ by \cite{KHP99}.

\section{Crossover of the Grassberger-A SCA model}

The PC conserving SCA by Grassberger is realized by the following
range-1 update rule 
(we show the configurations at $t-1$ and the probability $p$ of 
getting '$1$' at time $t$):
\begin{verbatim}
      t-1: 100 001 101 110 011 111 000 010
       t:   1   1   0  1-p 1-p  0   0   1
\end{verbatim}
The time evolution pattern in $1+1$ dimension, for small $p$ evolves 
towards a stripe-like ordered steady state (with double degeneration), 
while for $p > p_c = 0.1245(5)$ the kinks (the '$00$' and '$11$' 
pairs) survive. According to the classification of 
Wolfram~\cite{Wolfram} for $p=0$ we have the Rule-94, class 1 CA, 
while the $p=1$ limit is the chaotic Rule-22 deterministic CA.
This model has been investigated from damage spreading point of view
by us~\cite{odme98}.

\begin{figure}[ht]
\begin{center}
\epsfxsize=70mm
\epsffile{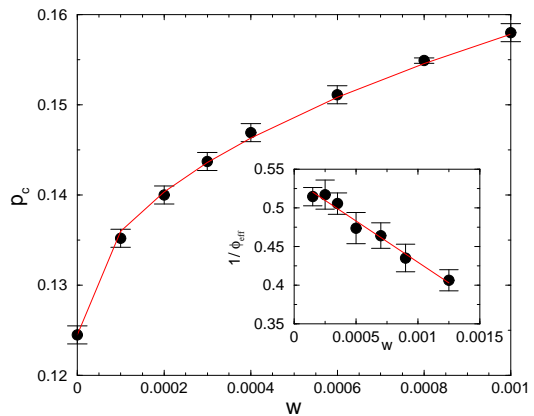}
\caption{(Color online) Critical point shift of the Grassberger's 
A SCA model to DP (bullets). The lines show power-law type 
fitting on the numerical data. The insert shows the local slopes
of the crossover exponent defined as (\ref{reff}).}
\label{Gfig}
\end{center}
\end{figure}

Now we extend this model in such a way that a $Z_2$ symmetry breaking
occurs in the absorbing state. This can be achieved by favoring one
of the absorbing phases ($'10101010'$ or $'01010101'$) 
shifted by a single site. Therefore we modified the transition rates 
\begin{verbatim}
      t-1: 100 001 101 110  011  111 000 010
       t:   1   1   0  1-p' 1-p'  0   0   1
\end{verbatim}
such that $p'=p+w$ at odd and $p'=p-w$ at even sites.
The simulations were run on $L=10^5$ sized rings up to
$t_{max}=10^6$ MCS. The location of the critical point
$p(w)$ is shown on Fig.~\ref{Gfig}.
To get more precise estimates for the crossover exponent
we determined the local slopes of the data points:
\begin{equation}
1/\phi_{eff}(w_i) = \frac {\ln r_c(w_i) - \ln r_c(w_{i-1})} 
{\ln(w_i) - \ln(w_{i-1})} \ \ .
\label{reff}
\end{equation}
This is plotted in the insert of Fig.\ref{Gfig} and in the
$w\to 0$ limit one can read-off the $1/\phi=0.53(2)$ linear
extrapolation value (with least squares error estimates)
in agreement with the value for NEKIMCA+h. As the figure
shows the correction to scaling can not be neglected.

\section{Crossover in a BARWe model}

To study the PC to DP crossover in an other way 
we take a simple version of a BARWe particle model 
introduced in \cite{ZAM03} and generalized in \cite{OdSzo} 
(ZAMb model) to allow phase transition at finite branching rate.
This model is defined on the one dimensional lattice as follows.
An occupied site is chosen randomly and is tried for diffusion, 
with probability $D$, or branching with probability 
$\sigma = 1 - D$); 
while the time is increased by $1/N$, where $N$ is the number of occupied
(active) sites. In a diffusion step the particle jumps to its
randomly chosen nearest neighbor sites. If the  
site is occupied both particles are annihilated 
with probability $r$. On the other hand the jump
is rejected with probability $1-r$. The branching
process involves the creation of two new particles
around the neighborhood. If either, or both
neighboring sites is previously occupied the target
site(s) become empty with probability $r$. Otherwise
the lattice remains unaltered with probability $1-r$.

\begin{figure}[ht]
\begin{center}
\epsfxsize=70mm
\epsffile{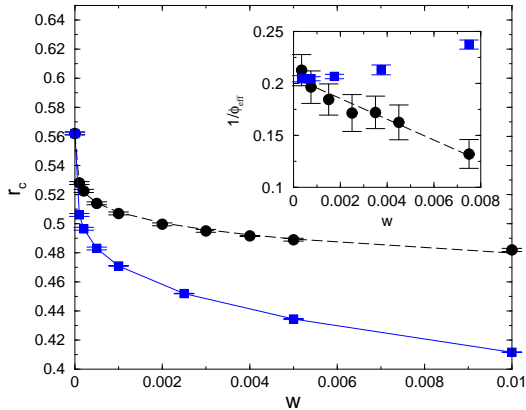}
\caption{(Color online) Critical point shift in the ZAMb model to DP
as the result of $AA\to A$ (bullets) and $A\to\emptyset$ (boxes)
parity conservation breaking. 
Dashed lines are power-law type fitting to numerical data. 
The insert shows the corresponding effective exponents 
defined as (\ref{reff}).}
\label{dcwfig}
\end{center}
\end{figure}

To induce a crossover to DP we introduced a spontaneous 
particle removal $A\to\emptyset$ or a coagulation $AA\to A$
with a small probability $w$ to the above reactions.
The simulations were done on lattices of size $L=10^5$ with
periodic boundary conditions for diffusion rate $D=0.2$
(the convergence of GMF approximations was found to be 
rather good for this diffusion rate in \cite{OdSzo}).
For $w=0$ we used the critical point value determined in 
\cite{OdSzo}: $r_c=0.562(1)$. As we increase $w$ the critical 
point $r_c(D)$ shifts as shown on Figure \ref{dcwfig}.
A simple power-law fitting for the data in the range $w\in (0,0.1)$ 
of the form \ref{crossscal} resulted in $\Delta r_c = 0.189 w^{0.181}$. 
A more precise estimate, which takes the corrections to scaling into 
account can be obtained by calculating the local slopes (\ref{reff}).
By plotting the effective exponent as the function of $w$ a linear
extrapolation resulted in $1/\phi = 0.205(5)$ both for the $AA\to A$ 
and the $A\to\emptyset$ parity breaking reactions 
\footnote{Note that the sign of correction to scaling is different
for the two crossover versions}.

\section{GMF+CAM calculations}

The simulation results were complemented by analytical cluster 
mean-field approximations and coherent anomaly extrapolations.
The generalized (cluster) mean-field method (GMF) suggested for
SCA by \cite{gut} and for dynamical RD models by \cite{dick}
has been shown to be successful approach for exploring the phase 
diagram of nonequilibrium models (see for example \cite{OSz05,Od1240} and
the references in \cite{Orev}).
It's extension with the coherent anomaly (CAM) extrapolation \cite{suzCAM}
enables one to extract the true scaling behavior.

In GMF we set up equations for the steady state of the
system based on $N$-point block probabilities. Correlations with
a range greater than $N$ are neglected.
By increasing $N$ from $1$ (traditional mean-field) step by step we
take into account more and more correlations and get better
approximations.
One can set up a master equations for the  $P_N$ block probabilities as
\begin{equation}
\frac{\partial P_N(\{s_i\})}{\partial t} = f\left (P_N(\{s_i\})\right) \ ,
\label{mastereq}
\end{equation}
where the site variables take the values: $s_i=\emptyset,A$.
During the solution of these equations one estimates larger than $N$
sized block probabilities by the maximum overlap approximation: 
\begin{equation}
P_{N+1}(s_1,...s_{N+1}) \simeq 
\frac{P_N(s_1,...s_{N}) P_N(s_2,...s_{N+1})}
{P_N(s_2,...s_{N},\emptyset)+P_N(s_2,...s_{N},A)}  \ . 
\label{approxeq}
\end{equation}
Taking into account the spatial symmetries and the conservation of
probability for the maximal, $N=9$ approximation of this work 
we had to find the solution of a set of nonlinear equations 
of 272 independent variables.

\begin{figure}[ht]
\begin{center}
\epsfxsize=70mm
\epsffile{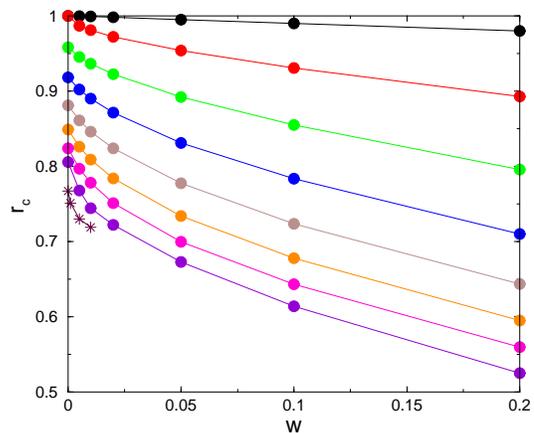}
\caption{(Color online) Critical point shift of the ZAMb model 
to DP (bullets) obtained by the GMF method for various cluster 
sizes ($N=1,2,...,9$ top to bottom).}
\label{dcwGMF}
\end{center}
\end{figure}

The steady state solutions of the $N$-cluster approximations
for the NEKIMCA ($1\le N\le 8$) and the ZAMb model ($1\le N\le 9$)
have been determined and the corresponding densities are calculated 
numerically. The phase transition points
are obtained for several values of the crossover parameter in both cases.
In case of the ZAMb model (at $D=0.2$) they are plotted on Fig.\ref{dcwGMF}
As one can see for $N=1$ (site mean-field) the critical point $r_c(N,w)$ 
exhibits linear relation without corrections to scaling. This also
true for the ZAMB+sink case. This means the $1/\phi_{MF}=1$ mean-field 
value for both kinds of crossover. 

For $N>1$ the  $r_c(N,w)$ curves pick up $w^{1/2}$ type of corrections 
to scaling (the leading order singularity is expected to remain
mean-field like). Assuming a scaling with the correction form
\begin{equation} 
r_c(N,w) = a(N) w + b(N) w^{1/2} \ , \label{camfitform} 
\end{equation}
which is natural if we swap the axes $r_c$ and $w$,
one can determine the amplitudes ($a(N)$) of the leading order term.
Following the envelope scaling hypothesis of \cite{suzCAM}
the Coherent Anomaly Method (CAM) extrapolation can be performed 
on the amplitude data in the $N\to\infty$ limit.

According to CAM the amplitudes $a(N)$ of the cluster mean-field
singularities in the leading order scale as 
\begin{equation}
|a(N)| \propto |r_c(N)-r_c|^{1/\phi - 1/\phi_{MF}} \label{anoscal}
\end{equation}
allowing to estimate the $1/\phi$ exponent of the true singular 
behavior (Eq.(\ref{crossscal})).
The $a(N)$ amplitudes were determined by a next leading order fitting 
form (\ref{camfitform}) on the the $r_c(N,w)$ data in the neighborhood 
of $r_c(N,0)$ for the ZAMb model.
In case of the NEKIMCA the same procedure has been applied for the
$\tilde\delta_c(N,h)$ crossover critical point GMF results.  

\begin{figure}
\begin{center}
\epsfxsize=70mm
\centerline{\epsffile{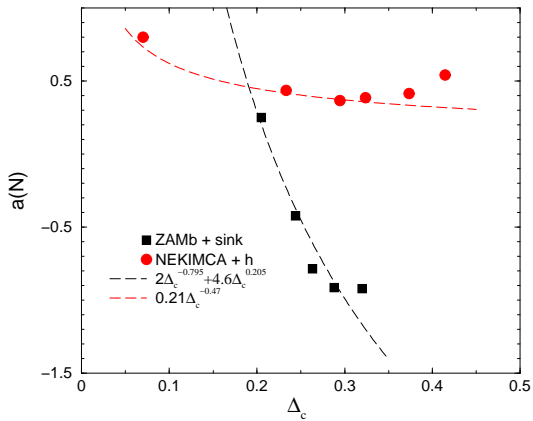}}
\caption{(Color online) CAM extrapolation results for the anomaly amplitudes 
$a(N)$. Bullets: NEKIMCA+h ($3\le N\le 8$ right to left), 
boxes: ZAMb + sink ($5\le N\le 9$ right to left)).
The numerical errors are smaller than the symbol sizes. 
The dashed lines show the expected critical behavior from simulations.}
\label{figcam}
\end{center}
\end{figure}

The highest $a(N)$ amplitudes with a CAM fitting form 
(which takes into account possible scaling corrections \cite{OSz05})
\begin{equation}
a(N) =  c \Delta_c^x + d \Delta_c^{x+1} \ , \label{aN}
\end{equation}
where $\Delta_c(N) = |\tilde\delta_c - \tilde\delta_c(N)|$ 
for NEKIMCA and $\Delta_c(N) = |r_c - r_c(N)|$ for the 
ZAMb are plotted on Fig.~\ref{figcam}.
In this form the nonuniversal fitting parameters are: $c$, $d$ and
the anomaly exponent is : $x=1/\phi - 1/\phi_{MF}$.
By the simulations (apart from the constants $c$ and $d$) 
we have an expected behavior for (\ref{aN}) both for ZAMb 
and NEKIMCA. 
This is plotted on Fig.~\ref{figcam} by the dashed lines.
As one can see our GMF data can be fitted with those lines,
(with the values: $c=2$, $d=4.6$ for ZAMb and 
$c=0.21$, $d=0$ for NEKIMCA) but the amplitudes for $N<10$ 
just start to converge towards the asymptotic scaling curves. 
This suggests that larger $N$-cluster
approximations should be determined. However the numerical 
instability of root finding of the GMF method (within the space 
of more than $500$ variable, nonlinear system) prevented us to 
go further.
Still the different behavior for NEKIMCA and ZAMb can be justified
by this figure.

\section{Conclusions and discussion}

\begin{figure}
\begin{center}
\epsfxsize=70mm
\centerline{\epsffile{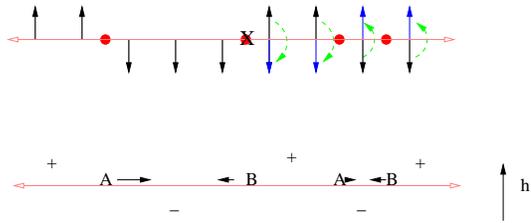}}
\caption{(Color online) Schematic mapping of the two different 
kinds of crossovers between the kink and the spin model. 
The top graph shows that the removal of a single kink corresponds 
to flipping of all spins to the right from position ``X''.
The bottom graph demonstrates that an external magnetic field
in the spin model causes repulsion/attraction force is among
domain walls (horizontal arrows) and an effectively two-component
reaction-diffusion system of kinks (particles) emerges.} 
\label{lekep}
\end{center}
\end{figure}

We have performed simulations and GMF+CAM approximations
for various versions of the PC class model crossovers towards 
the "DP class". By determining the crossover exponents we have found
and outstanding difference between the effect of a symmetry breaking 
field ($1/\phi=0.53(2)$) and the parity conservation breaking 
($1/\phi=0.205(5)$).
Although the NEKIMCA and the BARWe are dual models, 
the two types of crossover operators can't be mapped onto each 
other. Furthermore they can't be mapped onto a local one in the
dual system. The $A\to\emptyset$ process in the spin language
would mean flipping of all spins from a given site 
(see Fig.\ref{lekep}). Such transformation can't even be done
in finite systems with periodic boundary conditions.

A $Z_2$ symmetry breaking in the BARWe model favors the '$+$' (up)
or '$-$' (down) oriented spins depending its sign. In case of a '$+$'
preference (see Fig.\ref{lekep}) this means that '$+$'domains broaden 
on the expense of '$-$' domains, hence odd-even kinks 
(the corresponding particle pairs in BARWe) attract and even-odd 
pairs repel each other (shown by the horizontal arrows).
Therefore that particle model becomes an effectively two-component,
parity conserving one, and the sufficient conditions of the DP 
hypothesis do not hold.
Still one can see a DP type of decay in the global order parameter
because calling a '$+$' domain a macroscopic particle '$X$',
an effective DP process:  $X\to\emptyset$, $X\to 2X$, $2X\to X$ 
describes it's dynamics.

However there are certain operators which do not exhibit DP
type behavior. For example in case of cluster spreading it is easy
to see that there are two sectors in this model. An odd parity one,
with $\beta' = 0$ final survival probability 
defined as
\begin{equation}
\label{PScaling}
P_\infty \propto |p-p_c|^{\beta^\prime} \
\end{equation}
and an even parity one, with normal, $\beta'=\beta$ DP exponent. As a
consequence the hyper-scaling relation connecting the cluster 
spreading exponents and the rapidity reversal symmetry of DP is not
satisfied for this critical behavior. The universality class of
the DP is split into sub-classes. The subclass of models 
with BARWe dynamics with broken $Z_2$ symmetry (let's call it DP-2) 
is different from that of the ordinary 1+1 dimensional DP.
Since the difference is manifested in the cluster spreading behavior,
which is a consequence of a special initial condition, similarly to
the terminology used in equilibrium models with different surface
classes we do not claim the splitting of the DP class itself. 

This sub-class behavior is similar to the one, which is observed in the
pair contact process, where the frozen monomers cause different
cluster behavior from that of the simple contact process.
Also that kind of DP-2 behavior should arise, when one introduces 
spin-anisotropy in the NEKIM \cite{ujcikk,MeOd2003} and the effective
two-component BARWe model exhibits DP type of phase transition for
finite branching rate in the global order parameter, but different 
scaling behavior occurs in terms the cluster spreading.
Furthermore this model exhibits a re-entrant phase diagram and
at zero branching rate one finds another phase transition belonging
to the two-component BARWe model \cite{Cardy-Tauber}, but the cluster
spreading behavior, which is sensitive to the spin anisotropy is 
different again.
Therefore such sub-classes are not at all rare among 
low-dimensional reaction-diffusion models.

We would like to point out that by using an alternative definition for 
universality classes a different interpretation can also be given for
the same numerical results presented here.
Some authors define a university class by the field-theoretic
action of the model (without specifying the fixed point)
instead of the models exhibiting the same set of known
critical exponents as we do.
Since the action of BARWe particles \cite{Cardy-Tauber} 
and the action of the NEKIM spin model \cite{CCDDM05} 
(called GV model) do not agree, they should belong to 
different universality classes, which "intersect" in one 
dimension only \cite{Hayeof}. According to this picture the 
applied symmetry breaking to the PC and GV class models 
($Z_2$ breaking in GV and particle removal in BARWe) 
should end up in DP class somehow.

However we think that our definition of universality class,
and therefore our interpretation, is more precise.
It can describe the critical behavior of (particle) system 
for which no proper field theoretical action has been found.
Even within a field theory multiple fixed point solutions 
(corresponding to multiple critical points) can exist and
the relevancy of terms affecting the stability of a fixed 
point is not clear in many cases. 

Furthermore a coarse grained field theory may not capture
all scaling details of a particle model. In a discrete particle 
model diffusive annihilating particles can die out within 
finite time (even in an infinite sized system) due to the 
recurrence relation of random walks in one dimension 
\footnote{Note that this is not a finite size effect}. 
The particle survival probability decays asymptotically as 
$P_s(t)\propto t^{-\delta}$ and the final survival probability 
scales as (\ref{PScaling}). As a consequence a thermodynamic 
limit, in which the particle density are kept finite can't 
be established.

On the other hand in field theories of continuous variables
the survival probability is always unity and it had been unclear 
if the exponent $\delta$ and $\beta'$ could be defined sensibly. 
To overcome this discrepancy a reinterpretation of survival
probability was proposed in \cite{MGT}. 
This alternative definition of $\delta$ in field theory resulted 
in correct scaling exponents and hyper-scaling relations 
corresponding to symmetries. 

Since in our case the ``thermodynamic limit'' can't be
restricted to infinite particle number the cluster spreading 
behavior starting from finite number of particles is relevant 
from the universality point of view as in many papers
investigating absorbing phase transition via this approach
(see \cite{GrasTor,PCP,PCP2,DickMar,Hayeof,Orev,Svenrev}.
The cluster spreading behavior (the finite survival probability,
pair-connectedness functions, avalanche distributions ... etc. ) 
are sensitive to the initial parity of particles, hence strictly 
speaking the DP-2 subclass is not identical to the DP 
class in 1+1 dimension.

\vskip 0.5cm

\noindent
{\bf Acknowledgments:}\\
The authors thank Su-Chan Park and Hyunggyu Park for the useful communications.
Support from Hungarian research funds OTKA (Grant No. T046129) is acknowledged.
The authors thank the access to the HUNGRID and LCG-GRID.

\bibliography{ws-book9x6}

\end{document}